\documentclass[onecolumn,preprintnumbers,nofootinbib,amsmath,amssymb,superscriptaddress,11pt,a4paper]{article}
\usepackage{amsmath}
\usepackage{graphics}
\usepackage{hyperref}
\usepackage{amsfonts}
\usepackage{amssymb}%
\usepackage{enumitem}
\usepackage{geometry}
\usepackage{graphicx}% Include figure files
\usepackage{dcolumn}
\usepackage{setspace}
\providecommand{\U}[1]{\protect\rule{.1in}{.1in}}

\newtheorem{Prop}{Proposition}

\newtheorem{Ass'}{Assumption'}

\newcommand{\be}{\begin{equation}}
\newcommand{\ee}{\end{equation}}

\onehalfspace
\begin{document}
\title{Faster identification of faster Formula 1 drivers via time-rank duality}
\author{John Fry\footnote{Centre for Mathematical Sciences, School of Natural Sciences, University of Hull, Hull, HU6 7RX, UK.\newline
Email: J.M.Fry@hull.ac.uk},$\quad$ Tom Brighton\footnote{Centre for Mathematical Sciences, School of Natural Sciences, University of Hull, Hull, HU6 7RX, UK.\newline
Email: thomasbrighton02@gmail.com}$\quad$ and$\quad$ Silvio Fanzon\footnote{Centre for Mathematical Sciences, School of Natural Sciences, University of Hull, Hull, HU6 7RX, UK.\newline
Email: S.Fanzon@hull.ac.uk}} 
 \date{March 2024} \maketitle \setcounter{tocdepth}{2}
\begin{abstract}
Two natural ways of modelling Formula 1 race outcomes are a probabilistic approach, based on the exponential distribution, and econometric modelling of the ranks. Both approaches lead to exactly soluble race-winning probabilities. Equating race-winning probabilities leads to a set of equivalent parametrisations. This time-rank duality is attractive theoretically and leads to quicker ways of dis-entangling driver and car level effects.
\end{abstract}



% Codes used:
% 62P25 Applications of statistics to social sciences
% 91-10 Mathematical modeling or simulation for problems pertaining to game theory, economics, and finance
% 62M99 Inference from stochastic processes: None of the above, but in this section
% 65C05 Monte Carlo methods

%Codes worth considering:
% 62J05 Linear regression; mixed models
% 60-08 Computational methods for problems pertaining to probability theory
% 62M10 Time series, auto-correlation, regression, etc. in statistics (GARCH)

\section{Introduction}
Modelling Formula 1 races is an interesting econometric problem  (Bell et al., 2016; van Kesteren and Bergkamp, 2023)  of significant wider interest (Maurya, 2021). It is of interest to separate out driver-level and car-level effects. Previously, such an analysis has only been possible over longer time periods (Bell et al., 2016; Eichenberger and Stadelmann, 2009; van Kesteren and Bergkamp, 2023). Here, we present a solution that requires only one season of previous data.

Formula 1 races are most easily modelled assuming car finishing times are independently and exponentially distributed random variables. Under this assumption race-winning probabilities can be written down in closed from. This tractability also enables relatively easy model calibration via bookmakers' odds. However, this approach is at odds with much of the publicly-available race data. Final race-finishing times are typically unavailable as lapped cars do not typically finish the full race distance. In contrast, the most convenient way of modelling publicly-available race data is regression modelling of  the final finishing position (Eichenberger and Stadelmann, 2009).
Thus, in this paper, we combine both modes of analysis -- an approach we term time-rank duality.

The layout of this paper is as follows. Section 2 outlines a probabilistic approach to modelling race-finishing times and model calibration via bookmakers' odds. Section 3 establishes theoretical duality between this probabilistic approach and regression modelling of the final rank. Firstly, we show that a regression model for ranks can be used to estimate race-winning probabilities and then to an equivalent exponential-distribution parameterisation using the method in Section 2. Secondly, we show that, under the simplifying assumption of homoscedasticity, regression parameters can be reverse-engineered from a set of race-winning probabilities e.g. those corrsesponding to a given exponential-distribution parameterisation or a set of bookmakers' odds. Section 4 discusses empirical regression modelling of historical results. Section 5 combines both approaches to enable quicker identification of individual driver-level effects. Section 6 concludes. A mathematical appendix is contained at the end of the paper.

\section{Probabilistic approach to modelling finishing times}
Models based around the exponential distribution are amongst the most convenient ways to model Formula 1 races. This is due to its tractability alongside its usage in classical applied probability models. A related formulation based on the Weibull distribution is explored in the Appendix. Whilst its non-constant hazard function may be more physically realistic, the Weibull distribution may be more cumbersome in applications due to its additional shape parameter.

Suppose, for the sake of simplicity, that a race consists of $n$ cars whose finishing times $T_1, T_2, \ldots, T_n$ are independent exponential distributions with parameters $\lambda_1, \lambda_2, \ldots, \lambda_n$. Independence is a common simplifying assumption in sports models (Scarf et al., 2019) but may be difficult to justify empirically. A standard result in probability theory (Grimmett and Stirzaker, 2020) gives:

%\newpage

\begin{Prop}$\quad$
\begin{enumerate}[label=\roman*.]
\item If $T_1, T_2, \ldots, T_n$ are independent and exponentially distributed with parameters $\lambda_1, \lambda_2, \ldots, \lambda_n$ then
$$
\min\left\{T_1, T_2, \ldots, T_n\right\}\sim\exp\left(\sum_{i=1}^n\lambda_i\right).
$$
\item If $X$ and $Y$ are independent exponential distributions with parameters $\lambda_X$ and $\lambda_Y$ then 
\begin{eqnarray*}
Pr(X{\leq}Y)=\frac{\lambda_X}{\lambda_X+\lambda_Y}.
\end{eqnarray*}
\item Consider the Formula 1 race with independent and exponentially distributed finishing times as outlined above. Then 
\begin{eqnarray*}
    Pr(\mbox{Car $j$ wins})=\frac{\lambda_j}{\sum_{i=1}^n\lambda_i}.
\end{eqnarray*}
\end{enumerate}
\end{Prop}
%\textbf{Proof of Proposition 1}\\

Proposition 1 shows that given a sequence of win probabilities $p_1, p_2, \ldots, p_n$, calculated e.g. from bookmakers' odds, we can estimate the parameters $\lambda_i$. This can be done by minimising the Residual Sum of Squares (RSS):
\begin{eqnarray}
    \mbox{RSS}:=\sum_{i=1}^n\left(\frac{\lambda_i}{\lambda_1+\ldots+\lambda_n}-p_i\right)^2.\label{eq3}
\end{eqnarray}
The minimisation in (\ref{eq3}) can be done numerically. Results of the procedure applied to bookmakers' data are shown in Table 1. The R code and data to reproduce these results is openly available on GitHub\footnote{R codes and data files are available at \href{https://sfanzon.github.io/F1-Paper-Code}{https://sfanzon.github.io/F1-Paper-Code}.}. 
In Table 1 odds can be converted to probabilities as follows. The win probability corresponding to odds of 25/1 for Lewis Hamilton victory can be calculated via
\begin{eqnarray*}
    \frac{1-p}{p}=25;\ p=\frac{1}{26}.
\end{eqnarray*}
Win probabilities for the remaining drivers are calculated similarly, and then renormalised (\u{S}trumbelj, 2014) so that they sum to 1. These renormalised win probabilities are given in the fourth column of Table 1. Estimated $\hat\lambda$ values from the minimisation in (\ref{eq3}) are in the fifth column.

\begin{table}[h!]
\begin{center}
\begin{tabular}{|l|l|l|l|l|}
\hline
\textbf{Team} & \textbf{Car} & \textbf{Bookmakers} & \textbf{Implied} & \textbf{$\hat{\lambda}$} \\
& & \textbf{odds} & \textbf{win} &  \\
& & & \textbf{probability} &  \\
\hline
Mercedes & 	Lewis Hamilton & 	25/1 & 0.031655049 & 0.0081037902 \\
\hline
Mercedes &	George Russel &	25/1 & 0.031655049 & 0.0081037902 \\
 \hline
Red Bull	& Max Verstappen & 	2/9 & 0.673389233 &  0.1723897481 \\
\hline
Red Bull	&	Sergio Perez &	12/1 & 0.063310099 & 0.0162075831\\
 \hline
Ferrari & 	Charles Leclerc & 	25/1 & 0.031655049 & 0.0081037902 \\
\hline
Ferrari &  Carlos Sainz & 	28/1 & 0.028380389 & 0.0072654675\\
 \hline
Mclaren &	Lando Norris	& 12/1 & 0.063310099 & 0.0162075831 \\
\hline
Mclaren &	Oscar Piastri &	16/1 & 0.048413605 &  0.0123940343 \\
 \hline
Alpine & 	Estaban Ocon	& 500/1 & 0.001642777 & 0.0004205564 \\
\hline
Alpine &	Pierre Gasly &	500/1 & 0.001642777 & 0.0004205564\\
 \hline
Aston Martin &	Fernando Alonso & 80/1  & 0.01016088 & 0.0026012171 \\
\hline
Aston Martin &	Lance Stroll &	500/1 & 0.001642777 & 0.0004205564\\
 \hline
Haas &	Kevin Magnussen &	500/1 & 0.001642777 & 0.0004205564\\
\hline
Haas &	Nico Hulkenburg &	500/1 & 0.001642777 & 0.0004205564\\
\hline
Alfa Tauri &  Yuki Tsunoda & 	500/1 & 0.001642777 & 0.0004205564\\
\hline
Alfa Tauri &	Daniel Riccardo & 	500/1  & 0.001642777 & 0.0004205564\\
 \hline
Alfa Romeo &	Valterri Bottas	& 500/1 & 0.001642777 & 0.0004205564\\
\hline
Alfa Romeo &	Zhou Guanyu &	500/1 & 0.001642777 & 0.0004205564\\
 \hline
Williams &	Alex Albon & 	500/1 & 0.001642777 & 0.0004205564\\
\hline
Williams &	Logan Sergant & 	500/1 & 0.001642777 & 0.0004205564\\
\hline
\end{tabular}
\caption{Results of the model applied to betting data for the 2023 Qatar Grand Prix. (Source: \texttt{www.bet365.com}.)}\label{tabreg1}
\end{center}
\end{table}

\section{Econometric modelling of the final race ranking}
Empirical Formula 1 data are most commonly listed in terms of the rank rather than the strict finishing times. The analysis of historical race data is therefore most easily accomplished by regression modelling of the final rank obtained (Eichenberger and Stadelmann, 2009). This implicitly assumes a Gaussian model for sporting outcomes (Scarf et al., 2019).

Consider two related problems. Firstly, suppose that there are $n$ cars in the race and the final ranking $r_i$ of car $i$ can be approximated by a normal distribution: $r_i{\sim}N(\mu_i, \sigma^2_i)$. The approximate probability that car $i$ wins the race is given by 
\begin{eqnarray}
    p_i=Pr(r_i{\leq}1.5)=\Phi\left(\frac{1.5-\mu_i}{\sigma_i}\right),\label{eqrank1}
\end{eqnarray}
where $\Phi(\cdot)$ denotes the standard normal CDF. Secondly, suppose we are given a sequence of win probabilities $p_1, p_2, \ldots, p_n$ for Cars $1, 2, \ldots, n$. Under the simplifying assumption of $\sigma^2_i=\sigma^2$, equivalent to the classical normal linear regression model (Fry and Burke, 2022), from equation (\ref{eqrank1}) set
\begin{eqnarray}
    \Phi\left(\frac{1.5-\mu_i}{\sigma}\right)=p_i;\ \mu_i=1.5-\sigma\Phi^{-1}(p_i).\label{eqrank2}
\end{eqnarray}
Since the sum of the ranks is equal to $\frac{n(n+1)}{2}$ summing equation (\ref{eqrank2}) over $i$ gives
\begin{eqnarray}
    \frac{n(n+1)}{2}=1.5n-\sigma\sum_{i=1}^n\Phi^{-1}(p_i);\ \sigma=\frac{n-\frac{n^2}{2}}{\sum_{i=1}^n\Phi^{-1}(p_i)}.\label{eqrank3}
\end{eqnarray}
Combining equations (\ref{eqrank2}-\ref{eqrank3}) therefore gives the estimated $\mu_i$ values corresponding to the given win probabilities $p_i$. Table 2 applies this approach to estimate a set of $\hat{\mu}_i$ and $\hat{\sigma}^2$ regression parameters for the bookmakers' data in Table 1.

\begin{table}[h!]
\begin{center}
\begin{tabular}{|l|l|l|l|l|}
\hline
\textbf{Team} & \textbf{Car} & \textbf{Bookmakers} & \textbf{Implied} & \textbf{$\hat{\mu}_i$} \\
& & \textbf{odds} & \textbf{win} & \\
& & & \textbf{probability} & \\
\hline
Mercedes & 	Lewis Hamilton & 	25/1 & 0.031655049 & 8.704026 \\
\hline
Mercedes &	George Russel &	25/1 & 0.031655049 & 8.704026 \\
 \hline
Red Bull	& Max Verstappen & 	2/9 & 0.673389233 &  -0.242969 \\
\hline
Red Bull	&	Sergio Perez &	12/1 & 0.063310099 & 7.426002\\
 \hline
Ferrari & 	Charles Leclerc & 	25/1 & 0.031655049 & 8.704026 \\
\hline
Ferrari &  Carlos Sainz & 	28/1 & 0.028380389 & 8.890783\\
 \hline
Mclaren &	Lando Norris	& 12/1 & 0.063310099 & 7.426002 \\
\hline
Mclaren &	Oscar Piastri &	16/1 & 0.048413605 &  7.941444 \\
 \hline
Alpine & 	Estaban Ocon	& 500/1 & 0.001642777 & 12.904103 \\
\hline
Alpine &	Pierre Gasly &	500/1 & 0.001642777 & 12.904103\\
 \hline
Aston Martin &	Fernando Alonso & 80/1  & 0.01016088 & 10.501519 \\
\hline
Aston Martin &	Lance Stroll &	500/1 & 0.001642777 & 12.904103\\
 \hline
Haas &	Kevin Magnussen &	500/1 & 0.001642777 & 12.904103\\
\hline
Haas &	Nico Hulkenburg &	500/1 & 0.001642777 &  12.904103\\
\hline
Alfa Tauri &  Yuki Tsunoda & 	500/1 & 0.001642777 & 12.904103\\
\hline
Alfa Tauri &	Daniel Riccardo & 	500/1  & 0.001642777 & 12.904103\\
 \hline
Alfa Romeo &	Valterri Bottas	& 500/1 & 0.001642777 & 12.904103\\
\hline
Alfa Romeo &	Zhou Guanyu &	500/1 & 0.001642777 & 12.904103\\
 \hline
Williams &	Alex Albon & 	500/1 & 0.001642777 & 12.904103\\
\hline
Williams &	Logan Sergant & 	500/1 & 0.001642777 & 12.904103\\
\hline
\end{tabular}
\caption{Implied regression parameters corresponding to betting data for the 2023 Qatar Grand Prix ($\hat{\sigma}=3.879374$). (Source: \texttt{www.bet365.com}.)}\label{tabreg1}
\end{center}
\end{table}

\section{Regression modelling of historical results}
In this section we calibrate the model to historical results (observed race rankings) from the 2022 season which was the last fully-completed season at the time of writing. This follows a similar approach to modelling historical results in Fry et al. (2021). Following Eichengreen and Stadelmann (2009) we regress the finishing position against the dummy variables corresponding to each of the constructors. We then use stepwise regression (Fry and Burke, 2022) to automatically choose the best model. We constrain all models fitted to include a dummy variable indicating the teams' second (less-favoured) driver. Forwards and stepwise regression choose the same model indicated below in Table 3. In contrast, backward selection suggests a more complex model. However, an $F$-test, not reported, is non-significant suggesting the simpler model in Table 3 should suffice. Negative and significant parameters in Table 3 indicate more successful constructors with lower expected final finishing positions. 

\begin{table}[h!]
\begin{center}
\begin{tabular}{|l|l|l|l|l|}
\hline
\textbf{Coefficient} &
                 \textbf{Estimate} &  \textbf{Std. Error} &  \textbf{$t$-value} & \textbf{$p$-value}   \\
                 \hline
(Intercept)   &    13.8420  &   0.3794 &  36.484  & 0.000\\
\hline
Second driver &       0.2160 &    0.4056 &  0.533 &  0.5946\\
\hline
Red Bull &      -9.6500  &   0.7170 & -13.459 & 0.000\\
\hline
Mercedes  &   -8.2700  &   0.7170 & -11.534 & 0.000\\
\hline
Ferrari &     -7.6900 &    0.7170 & -10.725 & 0.000\\
\hline
Mclaren &     -3.5500  &   0.7170  & -4.951 &0.000\\
\hline
Alpine &       -3.5500 &    0.7170 & -4.951 & 0.000\\
\hline
Aston Martin &  -1.7900  &   0.7170 & -2.496  & 0.0129 \\ 
\hline
\end{tabular}
\caption{Stepwise regression results obtained (constrained to include driver order term). $R^2$ value=0.3914.}\label{tabreg1}
\end{center}
\end{table}

\section{Distilling driver-level effects}
From the regression output in Table 3 a 95\% confidence interval for the second driver term is 
\begin{eqnarray}
    \mbox{Second driver confidence interval}=(-0.581, 1.013).\label{eqrank4}
\end{eqnarray}
The upper value of 1.013 produced in (\ref{eqrank4}) is physically meaningful. Suppose race orderings are completely determined by the level of the car. In this case positions 1-2 would be occupied by the best car, positions 3-4 by drivers of the second best car, positions 5-6 by drivers of the third best car etc. A difference in the average ranking greater than 1 indicates that the quality of the leading driver is sufficient to be able to out-perform the next best car on the grid.

Thus combining equation (\ref{eqrank4}) with implied regression parameters in Table 2 a difference between two drivers of the same team bigger than 1.013 implies an extraordinary level of performance beyond the quality of the car. Comparing drivers in this way suggests two drivers Max Verstappen (Red Bull) and Fernando Alonso (Aston Martin) out-perform their respective cars. Past academic research has previously highlighted Verstappen's level of performance as historically significant (van Kesteren and Bergkamp, 2023).

\section{Conclusions}
 It is interesting to separate out driver-level and car-level effects in Formula 1 racing. Previously, such an analysis has only been possible over longer time periods (Bell et al., 2016; Eichenberger and Stadelmann, 2009; van Kesteren and Bergkamp, 2023). Here, we present a solution that requires only one season's worth of previous data. We combine a probabilistic approach based on the exponential distribution with econometric modelling of the ranks (Eichenberger and Stadelmann, 2009). Both approaches enable the race-winning probabilities to be exactly solved analytically. Equating race-winning probabilities means that both approaches can be seen as equivalent to each other. Results suggest that of the current crop of drivers Max Verstappen and Fernando Alonso out-perform the level of the car that they drive. Results match previous suggestions that Verstappen's performance level is historically significant (van Kesteren and Bergkamp, 2023). Future work will adjust the above models to account for cars that fail to finish races. Substantial interest in the analytical modelling of sports remains (Singh et al., 2023).  

\section*{Appendix: Mathematical proofs}
In Proposition 2 we consider race times to be independent Weibull distributions with common shape parameter $k$. This is a small technical extension of Proposition 1, where finishing times are exponential. We present the proof for Proposition 2 below,  noting that Proposition 1 is the special case of $k=1$ in Proposition 2.
\begin{Prop}$\quad$
\begin{enumerate}[label=\roman*.]
\item If $T_1, \ldots, T_n$ are independent and Weibull distributed with parameters $(\lambda_1, k),  \ldots, (\lambda_n, k)$ then
$$
\min\left\{T_1, T_2, \ldots, T_n\right\}\sim\mbox{\rm Weibull}\left(\sum_{i=1}^n\lambda_i, k\right).
$$
\item If $X\sim\mbox{Weibull}(\lambda_X, k)$ and $Y\sim\mbox{Weibull}(\lambda_Y, k)$ and $X$ and $Y$ are independent then 
\begin{eqnarray*}
Pr(X{\leq}Y)=\frac{\lambda_X}{\lambda_X+\lambda_Y}.
\end{eqnarray*}
\item Consider the Formula 1 race with independent and Weibull distributed finishing times as outlined above. Then 
\begin{eqnarray*}
    Pr(\mbox{Car $j$ wins})=\frac{\lambda_j}{\sum_{i=1}^n\lambda_i}.
\end{eqnarray*}
\end{enumerate}
\end{Prop}
\textbf{Proof of Proposition 2}
\begin{enumerate}[label=\roman*.]
\item  $Pr(T_i{\geq}x)=e^{-\lambda_i{x^k}}$. Since all the $T_i$ are independent
\begin{eqnarray*}
Pr(T_1{\geq}x, \ldots, T_n{\geq}x)  =  e^{-\lambda_1{x^k}} \ldots e^{-\lambda_n{x^k}}.    
\end{eqnarray*}
This gives
$$
Pr(\min\{T_1, \ldots, T_n\}\leq{x})=1-e^{-\left(\sum_{i=1}^n\lambda_i\right)x^k}.
$$
\item Since $f_X(x)=k\lambda_Xx^{k-1}e^{-\lambda_Xx^k}$ and $f_Y(y)=k\lambda_Yy^{k-1}e^{-\lambda_Yy^k}$
\begin{eqnarray*}
    Pr(X{\leq}Y) & = & \int_0^{\infty}\int_0^yk^2\lambda_X\lambda_Yx^{k-1}y^{k-1}e^{-\lambda_Xx^k}e^{-\lambda_Yy^k}dxdy\\
    & = & \int_0^{\infty}k\lambda_Yy^{k-1}e^{-\lambda_Yy^k}\left[-e^{-\lambda_Xx^k}\right]_0^ydy\\
    & = & \int_0^{\infty}k\lambda_Yy^{k-1}e^{-\lambda_Yy^k}dy-\int_0^{\infty}k\lambda_Yy^{k-1}e^{-(\lambda_X+\lambda_Y)y^k}dy\\
    & = & 1-\frac{k\lambda_Y}{k(\lambda_X+\lambda_Y)}=\frac{\lambda_X}{\lambda_X+\lambda_Y}.
\end{eqnarray*}
    \item For the sake of argument suppose $j=1$. Then
    $$Pr(\mbox{Car 1 wins})=Pr(T_1{\leq}\min\left\{T_2, T_3, \ldots, T_n\right\}).$$
    Now $T_1$ and $\min\left\{T_2, T_3, \ldots, T_n\right\}$ are independent with $$T_1{\sim}\mbox{Weibull}(\lambda_1, k);\ \min\left\{T_2, T_3, \ldots, T_n\right\}\sim\mbox{Weibull}\left(\sum_{i{\geq}2}\lambda_i, k\right).$$ Hence the result follows from part ii.
\end{enumerate}
$\hspace*{\fill}\Box \break$

\section*{Acknowledgements}
The authors would like to acknowledge helpful and supportive comments from an anonymous reviewer. The usual disclaimer applies.

\end{document}